\documentstyle[aps,pra,preprint]{revtex}
\begin{document}
\title{Exact Time Evolution and Master Equations for the Damped 
Harmonic Oscillator}
\author{Robert Karrlein and Hermann Grabert}
\address{Fakult\"{a}t f\"{u}r Physik der Albert--Ludwigs--Universit\"{a}t,
 Hermann-Herder-Str.~3,\\ D-79104 Freiburg, Germany}
\date{August 26, 1996}
\maketitle
\begin{abstract}
Using the exact path integral solution for the damped harmonic
oscillator it is shown that in general there does not exist an exact
dissipative Liouville operator describing the dynamics of the
oscillator for arbitrary initial bath preparations. Exact non-stationary
Liouville operators can be found only for particular preparations. 
Three physically meaningful examples are examined.
An exact new master equation is derived for thermal initial
conditions. Second, the Liouville operator governing the time-evolution of
equilibrium correlations is obtained. Third, factorizing initial
conditions  are studied. Additionally, one can show that there are
approximate Liouville operators independent of the initial preparation
describing  the long time dynamics under appropriate conditions. The
general form of these approximate master equations is derived
and the coefficients are determined for special cases of the bath
spectral density including the Ohmic, Drude and weak coupling cases.
The connection with earlier work is discussed.
\end{abstract}
\section{Introduction}
Recently the problem of the reduced dynamics of a quantum system in
contact with a reservoir has gained renewed interest 
\cite{pechukas,petruc,ford}.
While the foundations of quantum dissipative processes were layed
already in the sixties \cite{rev60}, this early work was mainly
concerned with weakly damped systems and has relied on the Born and Markov
approximations. In this conventional approach the dynamics of the
dissipative quantum system is described in terms of quantum master or
Langevin equations. Important progress beyond the limitations of the weak
coupling approach was made in the eighties, in particular path
integral techniques were shown to be powerful means to describe
quantum dissipative systems \cite{rev80}. This has lead to unexpected
new results, as for instance the non-exponential decay of correlation
functions in the low temperature range \cite{gwt,jung}, not directly
available within the conventional master equation approach.

The new insight gained also provided a
basis for profound critique \cite{pechukas,ford,Grabert82a,talk} of
concepts developed in the context of the quantum master equation
approach such as complete positivity or the quantum regression
theorem. However, the results of the path
integral approach were rarely used to derive improved master equations
valid in the range of strong damping and/or low temperatures. In this
paper we shall address this problem for the ubiquitous quantum
dissipative system, the damped harmonic oscillator. Based on available
exact results \cite{report} we derive new generalized
quantum master equations describing the exact relaxation of mean
values and the time evolution of equilibrium correlation functions.
Whenever appropriate, the new results will be confronted and compared
with earlier findings. 

More specifically, the article is organized as follows. In the
following section the microscopic model of a harmonic oscillator coupled to a
harmonic bath is introduced. The exact time evolution of the
reduced density matrix of the oscillator \cite{report} will briefly be
summarized. In section III we examine whether the exact time evolution
of the density matrix is compatible with a generalized quantum master
equation with a time dependent Liouville operator. It is shown that in
general there is no Liouville operator independent of the initial
preparation. In section IV we study specific types of
initial preparations. Exact time-dependent Liouville operators
describing the time evolution of thermal initial conditions and of
correlation functions are derived. In the classical limit these
Liouville operators are shown to reduce to the Adelman Fokker-Planck
operator \cite{adelman}. We also examine the case of factorizing
initial conditions. 

In section V we investigate particular types of
damping leading to a time evolution of the density
matrix governed approximately by a time-independent Liouville
operator. Earlier results by Talkner \cite{talk} and by
Haake and Reibold \cite{reibold} are recovered. We discuss in detail
the limit of a weakly damped oscillator and derive a generalization of
the Agarwal equation \cite{agar}. Additional time 
coarse-graining is shown to lead to the quantum-optical master
equation by Weidlich and Haake \cite{hw}. Finally in section VI we
present our conclusions.

\section{Model Hamiltonian and Exact Time Evolution}
\noindent
The standard microscopic model \cite{ullersma,zwanzig,cal} for the
damped harmonic oscillator starts out from a Hamiltonian $H$ composed of
the oscillator part
\begin{equation}
H_O={p^2\over 2M}+{M\over 2}\omega_0^2 q^2,
\end{equation}
the bath Hamiltonian
\begin{equation}
H_{\rm R}=\sum_{n=1}^N {p_n^2\over 2 m_n}+{m_n\over 2} \omega_n^2 x_n^2,
\end{equation}
and an interaction part
\begin{equation}
H_{\rm OR}=-q \sum_{n=1}^N c_n x_n +q^2 \sum_{n=1}^N {c_n^2\over 2 m_n
\omega_n^2}.\label{OR}
\end{equation}
This model describes an oscillator with the associated classical
equation of motion
\begin{equation}
M\ddot{q}(t)+\int_0^tds\gamma(t-s)\dot{q}(s)+\omega_0^2 q(t)=0
\end{equation}
where 
\begin{equation}
\gamma(t)={1\over M}\sum_{n=1}^N{c_n^2\over
m_n\omega_n^2}\cos(\omega_n t)
\end{equation}
is the damping kernel.
The dynamics of the density matrix of the entire system (oscillator
and bath) is given by
\begin{equation}
W(t)=\exp(-i H t/\hbar) W(0) \exp(i H t/\hbar),
\end{equation}
from where the reduced density matrix of the oscillator is obtained by
tracing out the bath modes
\begin{equation}
\rho(t)={\rm Tr}_{\rm R} W(t).
\end{equation}
The path integral technique allows for a complete solution to this
problem for a large class of initial states. Since the derivation is
expounded in \cite{report}, we merely state the result
\begin{equation}
\rho(q_f,q_f',t)=\int dq_i dq_i' d\bar{q} d\bar{q}' 
J(q_f,q_f',t,q_i,q_i',\bar{q},\bar{q}')\lambda(q_i,q_i',\bar{q},\bar{q}')
\label{result}
\end{equation}
where $\lambda(q_i,q_i',\bar{q},\bar{q}')$ is the so-called
preparation function defining the initial non-equilibrium state $W(0)$
as a modification of the equilibrium state $W_\beta=\exp(-\beta
H)/{\rm Tr}\exp(-\beta H)$ in the subspace
of the oscillator. This modification can be produced by operators
$O_j$, $O_j'$ that act upon the particle only leaving the reservoir
coordinates unchanged
\begin{equation}
W(0)=\sum_j O_j W_\beta O_j'\label{OO}.
\end{equation}
We then have
\begin{equation}
\langle q|W(0)|q'\rangle=\int d\bar{q}d\bar{q}'\lambda(q,q',\bar{q},\bar{q}')
\langle\bar{q}|W_\beta|\bar{q}'\rangle.\label{anfang}
\end{equation}
where
\begin{equation}
\lambda(q,q',\bar{q},\bar{q}')=\sum_j \langle q|O_j|\bar{q}\rangle
\langle \bar{q}'|O_j'|q'\rangle.
\end{equation}
Several relevant examples for $\lambda(q,q',\bar{q},\bar{q}')$ are
discussed in \cite{report} and special cases will be considered below.
$J(q_f,q_f',t,q_i,q_i',\bar{q},\bar{q}')$ is the propagating
function describing the time evolution of the reduced
density matrix. In terms of the transformed coordinates
$r=(q+q')/2,x=q-q'$, it reads
\begin{equation}
J(x_f,r_f,t,x_i,r_i,\bar{x},\bar{r})={1\over N(t)} \exp\left[{i\over\hbar}
\Sigma(x_f,r_f,t,x_i,r_i,\bar{x},\bar{r})\right]
\label{prop}
\end{equation}
where
\begin{eqnarray}
\Sigma(x_f,r_f,t,x_i,r_i,\bar{x},\bar{r})&=&
i\left({\hbar\bar{r}^2\over 2\langle q^2\rangle}+{\langle
p^2\rangle\over 2\hbar}\bar{x}^2\right)\nonumber\\
& &+M[x_f r_f f_1(t)+x_i r_i f_2(t)- x_i r_f f_3(t) - x_f r_i f_4(t)]
\nonumber\\
& & + {i\over 2}M\left[x_i^2 R^{++}(t)+2x_f x_i R^{+-}(t)+x_f^2
R^{--}(t)\right]\label{Sigmakl}\\
& &+M\left[\bar{r}(x_i C_1^+(t)+x_f C_1^-(t))-i\bar{x}(x_i C_2^+(t)+x_f
C_2^-(t))\right]\nonumber
\end{eqnarray}
and 
\begin{equation}
N(t) = {2\pi\hbar\over M|f_3(t)|}(2\pi\langle q^2\rangle)^{1/2}.
\end{equation}
The functions $f_i(t)$, $R^{\pm\pm}(t)$, and $C_i^{\pm}(t)$ are given in
terms of the symmetrized part $S(t)$ and the 
anti-symmetrized part $A(t)=-{\hbar\over 2 M} G(t)$ of
the equilibrium coordinate autocorrelation function 
$C(t)=\langle q(t)q\rangle=S(t)+i A(t)$ 
in the following way:
\begin{eqnarray}
f_1(t)&=&f_2(t)={\dot{G}(t)\over G(t)},\label{f12}\\
f_3(t)&=&{1\over G(t)},\label{f3}\\
f_4(t)&=&-\ddot{G}(t)+{\dot{G}(t)^2\over G(t)},\label{f4}\\
R^{++}(t)&=&{M\over\hbar}\left\{{\langle p^2 \rangle\over M^2}
   +{f_3(t)\over\langle q^2\rangle}\left[2\langle q^2\rangle\dot{S}(t)
   +f_3(t)\left(\langle q^2\rangle^2-S^2(t)\right)\right]\right\},\\
R^{+-}(t)&=& {M\over\hbar}
   \left\{\ddot{S}(t)-f_1(t)\dot{S}(t)-
   {f_3(t)\over\langle q^2\rangle}
   \left[\dot{S}(t)S(t)
   +f_1(t)\left(\langle q^2\rangle^2-S^2(t)\right)\right]\right\},\\
R^{--}(t)&=&{M\over\hbar}\left\{{\langle p^2 \rangle\over M^2}
   -{1\over\langle q^2\rangle}\left[\dot{S}(t)-f_1(t)S(t)\right]^2
   +\langle q^2\rangle f_1^2(t)\right\},\\
C_1^+(t)&=&f_3(t){S(t)\over\langle q^2\rangle}-f_1(t),\\
C_1^-(t)&=&{\dot{S}(t)\over\langle q^2\rangle}
    -f_1(t){S(t)\over\langle q^2\rangle}+f_4(t),\\
C_2^+(t)&=&{M\over\hbar}\left[{\langle p^2 \rangle\over M^2}+
   f_3(t)\dot{S}(t)\right],\\
C_2^-(t)&=&{M\over\hbar}
   \left[\ddot{S}(t)-f_1(t)\dot{S}(t)\right].\label{ende}
\end{eqnarray}
In the sequel we shall give results mainly in terms of $S(t)$ and
$G(t)$ which both remain finite in the classical limit while $A(t)$ vanishes.
The Laplace transforms of $G(t)$ and $S(t)$ read in terms of the
Laplace transform $\hat{\gamma}(z)$ of the damping kernel \cite{report,GtoG+}
\begin{eqnarray}
\hat{G}(z)&=&\left[z^2 + \omega_0^2 + z\hat{\gamma}(z)\right]^{-1},\nonumber\\
\hat{S}(z)&=&{1\over M\beta}\sum_{n=-\infty}^\infty {z\over\nu_n^2-z^2}
[\hat{G}(z)-\hat{G}(|\nu_n|)].\label{defSG}
\end{eqnarray} 
Finally,
\begin{eqnarray}
\langle q^2\rangle&=&S(0)={1\over M\beta}\sum_{n=-\infty}^\infty
[\omega_0^2+\nu_n^2+|\nu_n|\hat{\gamma}(|\nu_n|)]^{-1},\label{defpq}\\
\langle p^2 \rangle&=&-M^2\ddot{S}(0)={M\over\beta}\sum_{n=-\infty}^\infty
{\omega_0^2+|\nu_n|\hat{\gamma}(|\nu_n|)\over
\omega_0^2+\nu_n^2+|\nu_n|\hat{\gamma}(|\nu_n|)}\nonumber.
\end{eqnarray}
are the equilibrium variances.
\section{Master equation}
We now want to investigate the conditions under which the time
evolution of the reduced density matrix can be described by a
master equation
\begin{equation}
{\partial\over \partial
t}\rho(x,r,t)={\cal L}(x,r,{\partial\over\partial x},{\partial \over
\partial r},t)\rho(x,r,t) 
\label{spatial}
\end{equation}
with a generally time-dependent Liouville operator ${\cal L}$.
Since $\Sigma$ is quadratic in $x$ and $r$, it is sufficient
to make the ansatz
\begin{eqnarray}
{\cal L} &=&L_c(t)+{\hbar\over M}
\left({\partial\over\partial x},{\partial \over \partial r}\right)
\left(\begin{array}{cc}X_{xx}(t) & X_{xr}(t)\\
                          X_{rx}(t) & X_{rr}(t)
                        \end{array}\right)
\left(\begin{array}{c}{\partial\over\partial x}\\{\partial\over\partial r}
      \end{array}\right)+\biggl(x,r\biggr)\left(\begin{array}{cc}
                         Y_{xx}(t) & Y_{xr}(t)\\
                         Y_{rx}(t) & Y_{rr}(t)
                        \end{array}\right)
\left(\begin{array}{c}{\partial\over\partial x}\\{\partial\over\partial r}
      \end{array}\right)\nonumber\\
& & -{M\over\hbar} \biggl(x,r\biggr)\left(\begin{array}{cc}
                         Z_{xx}(t) & Z_{xr}(t)\\
                         Z_{rx}(t) & Z_{rr}(t)
                        \end{array}\right)
\left(\begin{array}{c}x\\r
      \end{array}\right)
\label{matr}
\end{eqnarray}
with complex parameter $L_c(t)$ and complex matrices 
${\bf X}(t)$, ${\bf Y}(t)$ and ${\bf Z}(t)$. Without loss of
generality we may put $X_{rx}(t)=X_{xr}(t)$ and $Z_{rx}(t)=Z_{xr}(t)$.
Hence, there are eleven time-dependent functions in the ansatz (\ref{matr}). 
Using (\ref{result}), (\ref{prop}),(\ref{Sigmakl}), and
(\ref{matr}) to write the left hand and right hand sides of (\ref{spatial})
in explicit form, we 
find that a master equation exists for arbitrary preparation functions
provided that 15 ordinary differential equations are fulfilled.
Three of these, obtained by comparing the coefficients 
of $r_f^2,r_fr_i,r_i^2$, read
\begin{eqnarray}
0 & = & Z_{rr}(t) + X_{xx}(t) f_1(t)^{2} -i Y_{rx}(t) f_1(t), \nonumber\\
0 & = & -i Y_{rx}(t) + 2 X_{xx}(t)f_1(t),  \nonumber\\
0 & = & X_{xx}(t) f_4(t)^2.\nonumber
\end{eqnarray}
Inserting the solution $X_{xx}(t)=Y_{rx}(t)=Z_{rr}(t)=0$ into the 12 remaining
equations, one finds that the conditions obtained by comparing the
coefficients of $1$ and $r_fx_i$ imply $L_c(t)=Y_{rr}(t)$. 
Taking this into account, we are left with the following set of 11 equations: 
\begin{eqnarray}
\dot{f_1}(t) &=& 2i X_{xr}(t)f_1(t)^2+[Y_{rr}(t)+Y_{xx}(t)]
 f_1(t)+2i Z_{xr}(t)\label{dgl1}\\
\dot{f_2}(t) &=& 2i X_{xr}(t)f_4(t)f_3(t) \label{dgl2}\\
\dot{f_3}(t) &=& \left[Y_{rr}(t)+ 2i X_{xr}(t)
f_1(t)\right]f_3(t)\label{dgl3}\\
\dot{f_4}(t) &=& \left[Y_{xx}(t)+2i X_{xr}(t) 
f_1(t)\right]f_4(t)\label{dgl4}\\
\dot{R}^{++}(t)&=&-4i X_{xr}(t)f_3(t)R^{+-}(t)+
2X_{rr}(t)f_3^2(t)\label{dgl5}\\
\dot{R}^{+-}(t) &=& \left[Y_{xx}(t)+2i X_{xr}(t)f_1(t)\right]
  R^{+-}(t)\nonumber\\
  & & +\left[-2 X_{rr}(t)f_1(t)-2i
  X_{xr}(t)R^{--}(t)+iY_{xr}(t)\right]f_3(t)\label{dgl6}\\
\dot{R}^{--}(t)&=&2\left[Y_{xx}(t)+ 2i X_{xr}(t) 
f_1(t)\right] R^{--}(t)\nonumber\\
  & & +2Z_{xx}(t)-2i Y_{xr}(t)f_1(t)+2X_{rr}(t)f_1^2(t)
     \label{dgl7}\\
\dot{C}_1^+(t)&=&-2i X_{xr}(t)f_3(t) C_1^-(t)\label{c1+}\\
\dot{C}_1^-(t)&=&\left[Y_{xx}(t)+2i X_{xr}(t)f_1(t)\right]
C_1^-(t)\label{c1-}\\
\dot{C}_2^+(t)&=&-2i X_{xr}(t)f_3(t) C_2^-(t)\label{c2+}\\
\dot{C}_2^-(t)&=&\left[Y_{xx}(t)+2iX_{xr}(t)f_1(t)\right]C_2^-(t)\label{c2-}
\end{eqnarray}
Since only seven functions of the ansatz (\ref{matr}) remain to be
determined, the set (\ref{dgl1})-(\ref{c2-}) will be seen to have no
solution in general. To demonstrate this explicitly, let us first
disregard the equations (\ref{c1+})-(\ref{c2-}) which stem from comparing
coefficients involving the coordinates $\bar{x}$ and $\bar{r}$ of the 
preparation function. The remaining set of equations
(\ref{dgl1})-(\ref{dgl7}) has a unique solution and, the resulting
Liouville operator can be written in the form
\begin{equation}
{\cal L}(t)={i\hbar\over M}{\partial^2\over\partial x\partial r}
   -{iM\over\hbar}\gamma_{q}(t)rx-\gamma_{p}(t)x{\partial\over\partial x}
   -{iM\over \hbar}D_{q}(t)x{\partial\over\partial r}
   -{M^2\over\hbar^2}D_{p}(t)x^2.
\label{master}
\end{equation}
To see this one first notes that (\ref{dgl2}), (\ref{dgl3}), and (\ref{dgl5})
give 
\begin{equation}
X_{rr}(t)=Y_{rr}(t)=0\quad \mbox{and}\quad X_{xr}(t)=i/2.
\label{ais1}
\end{equation}
The remaining four
functions are then readily determined. For later convenience they are
expressed in terms of the four functions introduced in (\ref{master})
which are given by
\begin{eqnarray}
\gamma_{q}(t)&=&-2iZ_{xr}(t)=
   {\ddot{G}^2(t)-\dot{G}(t)\overdots{G}(t)\over \dot{G}^2(t)-
   G(t)\ddot{G}(t)},\nonumber\\
 \gamma_{p}(t)&=&-Y_{xx}(t)={G(t)\overdots{G}(t)-
   \dot{G}(t)\ddot{G}(t)\over \dot{G}^2(t)-
   G(t)\ddot{G}(t)},\label{coeffs}\\
D_{q}(t)&=&{i\hbar\over M} Y_{xr}(t)
   =\gamma_{q}(t)\langle q^2\rangle-{\langle p^2\rangle\over M^2}
   +{S(t)[\gamma_{p}(t)X(t)+\dot{X}(t)]\over \langle q^2\rangle G(t)}
   +\gamma_{p}(t)Y(t)+\dot{Y}(t),\nonumber\\
D_{p}(t)&=&{\hbar\over M}Z_{xx}(t)
   =\gamma_{p}(t){\langle p^2\rangle\over M^2}
   +{\dot{S}(t)[\gamma_{p}(t)X(t)+\dot{X}(t)]\over 
   \langle q^2\rangle G(t)}
   +{\dot{G}(t)[\gamma_{p}(t) Y(t)+\dot{Y}(t)]\over G(t)},\nonumber
\end{eqnarray}
where we have introduced
\begin{eqnarray}
X(t)&=&\dot{G}(t)S(t)-G(t)\dot{S}(t),\nonumber\\
Y(t)&=&G(t)\ddot{S}(t)-\dot{G}(t)\dot{S}(t).
\end{eqnarray}
Inserting $X_{xr}(t)=i/2$ and $Y_{rr}(t)=0$ into
(\ref{c1+})-(\ref{c2-}), it is readily seen that
(\ref{c1+}) and (\ref{c2+}) are already satisfied. However,
(\ref{c1-}) and (\ref{c2-}) are only fulfilled provided that
\begin{eqnarray}
\ddot{S}(t)+\gamma_{p}(t)\dot{S}(t)+\gamma_{q}(t)S(t) &=& 0\label{I}, \\
\overdots{S}(t)+\gamma_{p}(t)\ddot{S}(t)+\gamma_{q}(t)\dot{S}(t) &=& 0\label{II}.
\end{eqnarray}
Differentiating (\ref{I}) and subtracting (\ref{II}) we obtain
\begin{equation}
\dot{\gamma}_{p}(t)\dot{S}(t)+\dot{\gamma}_{q}(t)S(t)=0.\label{I-II}
\end{equation}
Now for $\dot{\gamma}_{p}(t)\neq 0$ this gives
\begin{equation}
{\dot{S}(t)\over S(t)}=-{\dot{\gamma}_{q}(t)\over
\dot{\gamma}_{p}(t)}={\dot{G}(t)\over G(t)}
\end{equation}
where the last equation follows by means of
(\ref{coeffs}). This implies
\begin{equation}
S(t)=c\, G(t)\quad  (t\ge 0) \label{bedingung}
\end{equation}
where $c$ is a real constant. Clearly, this condition, which is
equivalent to Onsager's regression hypothesis \cite{ford}, is never met
exactly whatever the form of the damping kernel. To see this
explicitly we note that the Taylor series 
of $S(t)$ and $G(t)$ start according to
\begin{eqnarray}
S(t)&=&\langle q^2\rangle-{\langle p^2\rangle\over 2M^2}t^2+{\cal O}
(t^4)\nonumber\\
G(t)&=&t+{\cal O}(t^3)
\label{entwick}
\end{eqnarray}
On the other hand, in the case $\dot{\gamma}_{p}(t)=0$, the condition 
(\ref{I-II}) is only fulfilled if $ \gamma_{p}(t)$ and $\gamma_{q}(t)$
are both independent of time. Further $S(t)$ must be of the form
\begin{equation}
S(t)=d_1 e^{-\lambda_1 t}+d_2 e^{-\lambda_2 t}\quad (t\ge 0),
\label{ohmic1}
\end{equation}
where $d_1$ and $d_2$ are complex constants and
$\lambda_{1/2}= \gamma_{p}/2\pm i\sqrt{\gamma_{q}-\gamma_{p}^2/4}$. 
Moreover, from
(\ref{coeffs}) we see that $ \gamma_{p}(t)$ is constant provided that $G(t)$ 
is of the form
\begin{equation}
G(t)=c_1 e^{-\lambda_1 t}+c_2 e^{-\lambda_2 t}\quad (t\ge 0),
\label{ohmic2}
\end{equation}
where $c_1$ and $c_2$ are complex constants.
It is now easily seen that (\ref{ohmic1}) and (\ref{ohmic2}) never
hold exactly except for $\gamma_{p}=0$, which means in the absence of damping.
Hence there is no {\em exact} master equation for the damped harmonic
oscillator with a Liouville operator ${\cal L}$ {\em independent} of the
preparation function.

In the remainder of this work we first consider specific
initial preparations for which the time evolution is described exactly
by a time-dependent Liouville operator. We then give examples for 
approximate Liouville operators valid for particular types of damping.

\section{Liouville operators for special initial preparations}
In the previous section we have shown that there is no exact
Liouville operator which is independent of the preparation
function. However, for certain preparations the set of equations
(\ref{dgl1})-(\ref{c2-}) which determine the time-dependend parameters in 
${\cal L}$ can be reduced allowing for an exact solution. In the
sequel three types of initial states will be considered.
 
\subsection{Thermal initial condition}
Let us first consider a system that is initially in a state
\begin{equation}
\langle q| W(0)|q'\rangle=r(q,q')\langle q|W_\beta|q'\rangle\label{initially}
\end{equation}
where $r(q,q')$ is an arbitrary function of $q$ and $q'$. This initial
condition allows in (\ref{OO}) only for operators
$O_j$ and $O_j'$ that are diagonal in position space.
It can be used to describe initial states resulting from position
measurements but excludes measurements of velocities or variables that
couple to the position and the momentum. Following Hakim and
Ambegaokar \cite{hakim} we call (\ref{initially}) a
{\em thermal initial condition}. Inspection of 
(\ref{anfang}) shows that the corresponding preparation function
is given by
\begin{equation}
\lambda(x_i,r_i,\bar{x},\bar{r})=r(x_i,r_i)\delta(\bar{x}-x_i)
\delta(\bar{r}-r_i).
\end{equation}
This form of the preparation function has the consequence that only
the difference of (\ref{c1-}) and (\ref{dgl4}) and likewise the
difference of (\ref{c2-}) and (\ref{dgl6}) must be fulfilled. Since 
(\ref{c1+}) and (\ref{c2+}) are again satisfied as a consequence of
(\ref{ais1}), there is indeed an exact solution
\begin{eqnarray}
\gamma_{q}(t)&=&{\dot{G}(t)\ddot{S}(t)-\ddot{G}(t)\dot{S}(t)
   \over\dot{G}(t)S(t)-G(t)\dot{S}(t)}\nonumber\\
\gamma_{p}(t)&=&{G(t) \ddot{S}(t)-\ddot{G}(t)S(t)
   \over\dot{G}(t)S(t)-G(t)\dot{S}(t)}\nonumber\\
D_{q}(t)&=&\gamma_{q}(t) \langle q^2\rangle-
   {\langle p^2\rangle\over M^2}\label{constrained}\\
D_{p}(t)&=&{\langle p^2\rangle\over M^2} \gamma_{p}(t).\nonumber
\end{eqnarray}
When these coefficients are inserted into (\ref{master}), we obtain a 
time-dependent Liouville operator valid for a large class of initial
states. Using the relations
\begin{eqnarray}
x&\to&[q,\cdot\ ],\ r\to{1\over 2}\{q, \cdot\ \}, \nonumber\\
{\partial\over\partial x}&\to&{i\over 2\hbar}\{p,\cdot\ \},\ 
{\partial\over\partial r}\to{i\over \hbar}[p,\cdot\ ]\label{rules}
\end{eqnarray}
the resulting exact master equation can be written in the form
\begin{eqnarray}
\dot{\rho}(t)&=&-{iM\over\hbar}\gamma_{q}(t)\left[q,{1\over 2}\{q,\rho(t)\}
   +{i\over\hbar}\langle q^2\rangle[p,\rho(t)]\right]\nonumber\\
& &-{i\over\hbar}\gamma_{p}(t)\left[q,{1\over 2}\{p,\rho(t)\}
   -{i\over\hbar}\langle p^2\rangle[q,\rho(t)]\right]\label{neu}\\
& &-{i\over M\hbar}\left[p,{1\over 2}\{p,\rho(t)\}
   -{i\over\hbar}\langle p^2\rangle[q,\rho(t)]\right]\nonumber.
\end{eqnarray}
Here $[A,B]=AB-BA$ denotes the commutator and $\{A,B\}=AB+BA$ the
anti-commutator.

To see the connection of this new master equation with earlier
results, we rewrite it in terms of the Wigner transform of the
reduced density matrix defined by
\begin{equation}
w(p,q,t)=\int dx \exp\left(-{i\over\hbar}xp\right) \rho(x,q,t).
\end{equation}
Using the rules 
\begin{eqnarray}
[q,\cdot\ ]&\to -{\hbar\over i}{\partial\over\partial p}&,\ 
\{q,\cdot\  \}\to 2q\\
{[p,\cdot\ ]} &\to {\hbar\over i}{\partial\over\partial q}&,\ 
\{p,\cdot\ \}\to 2p
\end{eqnarray}
we find from (\ref{neu})
\begin{equation}
\dot{w}(p,q,t)=\left\{
{\partial\over\partial p}M\gamma_{q}(t)
\left[q+{\partial\over\partial q}\langle q^2\rangle\right]
+\left[{\partial\over\partial p}\gamma_{p}(t)
-{\partial\over\partial q}{1\over M}\right]
\left[p+{\partial\over\partial p}\langle p^2\rangle\right]
\right\}w(p,q,t).\label{wigner2}
\end{equation}
This is of the form of a generalized
Fokker-Planck equation. For the classical harmonic oscillator an
equation of similar form was found by Adelman \cite{adelman} based on
the generalized classical Langevin equation 
\begin{equation}
\ddot{q}(t)+\omega_0^2q(t)+\int_0^tds\gamma(t-s)\dot{q}(s)={1\over M}\xi(t)
\label{gle}
\end{equation}
with a noise force $\xi(t)$ satisfying
\begin{equation}
\langle\xi(t)\rangle=0
\end{equation}
and
\begin{equation}
\langle \xi(t)\xi(0)\rangle=Mk_BT\gamma(t).
\end{equation}
The time-dependent coefficients of the Adelman equation are given by 
\begin{eqnarray}
\gamma_{q}(t)&=&{\ddot{C}_{\rm cl}^2(t)
   -\dot{C}_{\rm cl}(t)\overdots{C}_{\rm cl}(t)
   \over \dot{C}_{\rm cl}^2(t)-
   C_{\rm cl}(t)\ddot{C}_{\rm cl}(t)}\nonumber\\
\gamma_{p}(t)&=&{C_{\rm cl}(t)\overdots{C}_{\rm cl}(t)
   -\dot{C}_{\rm cl}(t) \ddot{C}_{\rm cl}(t)
   \over \dot{C}_{\rm cl}^2(t)
   -C_{\rm cl}(t)\ddot{C}_{\rm cl}(t)}\label{adel}
\end{eqnarray}
Since in the classical limit $S(t)$ reduces to $C_{\rm cl}(t)$, and
\begin{equation}
G(t)=-{M\over k_B T}\dot{C}_{\rm cl}(t)
\end{equation}
the coefficients (\ref{constrained}) reduce to (\ref{adel}). 
This shows that we have derived an exact quantum mechanical
generalization of the Adelman equation.
\subsection{Liouville operator for the time evolution of equilibrium
correlations}
In this section we investigate the time evolution of equilibrium correlation
functions 
\begin{equation}
\langle A(t) B\rangle ={\rm Tr}\left(A e^{-iHt/\hbar}BW_\beta
e^{iHt/\hbar}\right).
\end{equation}
where $A$ and $B$ are variables of the oscillator and Tr denotes the
trace over the Hilbert space of the entire system (oscillator and bath).
Inserting three partitions of unity and using 
\begin{equation}
{\rm Tr}=\int dq\, {\rm Tr_R}\langle q|\ \cdot\ |q\rangle,\label{corrfk}
\end{equation}
where ${\rm Tr_R}$ is the trace over the Hilbert space of the bath, 
(\ref{corrfk}) can be written as a fourfold integral
\begin{equation}
\langle A(t) B\rangle=\int dq_1dq_2dq_3dq_4 A(q_2,q_1)B(q_3,q_4)
P(q_1,q_2,t,q_3,q_4).\label{Pdarst}
\end{equation}
where $A(q,q')=\langle q|A|q'\rangle$ and
\begin{equation}
P(q_1,q_2,t,q_3,q_4)={\rm Tr_R}\left(\langle q_1|e^{-iHt/\hbar}| q_3\rangle
\langle q_4|W_\beta e^{iHt/\hbar}|q_2\rangle\right).
\end{equation}
This function contains complete information
about equilibrium correlation functions. 

$P(q_1,q_2,t,q_3,q_4)$ satisfies an exact master equation. To see this 
we first note that an equilibrium correlation function may be
calculated in the following way \cite{report}. One propagates the
initial reduced ``density matrix'' $\rho_B(0)=B \rho_\beta$ and takes
the expectation value of $A$ after time $t$. Hence,
\begin{equation}
\langle A(t) B\rangle={\rm tr}\left[A\rho_B(t)\right]=\int dq_fdq_f' 
A(q_f',q_f)\rho_B(q_f,q_f',t)
\end{equation}
where tr denotes the trace over the Hilbert space of the oscillator. 
Since the initial reduced ``density matrix'' $B\rho_\beta$ corresponds to the
``density matrix'' $BW_\beta$ of the entire system, the preparation
function reads
\begin{equation}
\lambda_B(q_i,q_i',\bar{q},\bar{q}')=B(q_i,\bar{q})\delta(\bar{q}'-q_i').
\label{prepB}
\end{equation}
Now, using 
\begin{equation}
\rho_B(q_f,q_f',t)=\int dq_idq_i'd\bar{q}d\bar{q}'J(q_f,q_f',t,q_i,q_i',
\bar{q},\bar{q}')\lambda_B(q_i,q_i',\bar{q},\bar{q}')
\end{equation}
we arrive at
\begin{equation}
\langle A(t) B\rangle=\int dq_fdq_f'dq_idq_i'd\bar{q}d\bar{q}'
A(q_f,q_f')B(q_i,\bar{q})J(q_f,q_f',t,q_i,q_i',\bar{q},\bar{q}')
\delta(\bar{q}'-q_i')
\end{equation}
which yields by comparison with (\ref{Pdarst})
\begin{equation}
P(q_1,q_2,t,q_3,q_4)=\int dy J(q_1,q_2,t,q_3,y,q_4,y).
\end{equation}
Inserting the explicit form of the propagating function and after the
Gaussian integration over $y$, we are left with an expression
containing $S(t)$ and $G(t)$ only in the combination
$S(t)-i\hbar G(t)/2M =C(t)$. One finds
\begin{eqnarray}
\lefteqn{P(x_f,r_f,t,x_i,r_i)={1\over \sqrt{2\pi N(t)}}\exp\Biggl\{{i M \over 2
\hbar}\biggl[
{\dot{N}(t)\over N(t)}(x_fr_f-x_ir_i)
+{\dot{N}(t)\langle q^2\rangle\over N(t)C(t)}(x_ir_f-x_f r_i)}\nonumber\\
& &-{iM\over\hbar}
\left[-{\langle p^2\rangle\over M^2}+{\dot{N}(t)^2\langle q^2\rangle 
\over 4 C(t)^2N(t)}\right](x_f^2+x_i^2)
-{iM\over\hbar}\left({\ddot{N}(t)\over C(t)}+{\dot{N}^2(t)\langle q^2\rangle 
\over 2 N(t) C^3(t)}\right)x_f x_i\\
& &+{i\hbar\over M}{\langle q^2\rangle\over N(t)}(r_f^2+r_i^2)
-2{i \hbar\over M}{C(t)\over N(t)}r_ir_f
\biggr]\Biggr\},\nonumber
\end{eqnarray}
where $r_f=(q_1+q_2)/2$, $x_f=q_1-q_2$, $r_i=(q_3+q_4)/2$, and $x_i=q_4-q_3$.
Further, we have introduced $N(t)=\langle q^2\rangle^2-C(t)^2$.
In view of the $\delta$ function in (\ref{prepB}) there are again
fewer conditions that must be satisfied by the time-dependent
coefficients of the Liouville operator. In fact, it is easily seen that 
\begin{equation}
\dot{P}(x_f,r_f,t,x_i,r_i)={\cal L}(x_f,r_f,{\partial\over\partial
x_f},{\partial \over \partial r_f},t) P(x_f,r_f,t,x_i,r_i),\label{Liou}
\end{equation}
with a Liouville operator ${\cal L}$ of the form (\ref{master}) with
the coefficients
\begin{eqnarray}
\gamma_{q}(t)&=&{\ddot{C}^2(t)-\dot{C}(t)\overdots{C}(t)\over 
\dot{C}^2(t)-C(t)\ddot{C}(t)}\nonumber\\
\gamma_{p}(t)&=&{C(t) \overdots{C}(t)-\dot{C}(t)\ddot{C}(t)
\over \dot{C}^2(t)-C(t)\ddot{C}(t)}\label{schramm}\\
D_{q}(t)&=&\gamma_{q}(t) \langle q^2\rangle-
{\langle p^2\rangle\over M^2}
\nonumber\\
D_{p}(t)&=&{\langle p^2\rangle\over M^2} \gamma_{p}(t).\nonumber
\end{eqnarray}
Hence $P(x_f,r_f,t,x_i,r_i)$ satisfies the exact evolution equation
\begin{eqnarray}
\dot{P}(x_f,r_f,t,x_i,r_i)&=&\Biggl[-{iM\over\hbar}\gamma_q(t)
x\left(r+\langle q^2\rangle{\partial \over\partial r}\right)\nonumber\\
&&-\left(\gamma_p(t)x+{\hbar\over iM}{\partial\over\partial
r}\right)\left({\partial\over\partial x}+{\langle p^2\rangle\over\hbar^2}
x\right)\Biggr]P(x_f,r_f,t,x_i,r_i).\label{evol}
\end{eqnarray}
To illuminate the virtue of this equation we note that it can be used
to calculate correlation functions in a quasi-classical manner.
Introducing the double Wigner transform
\begin{equation}
\tilde{P}(p_f,q_f,t,p_i,q_i)={1\over(2\pi\hbar)^2}\int dx_fdx_i
P(x_f,q_f,t,x_i,q_i) \exp\left[{i\over\hbar}(-x_fp_f+x_ip_i)\right].
\end{equation}
and the Wigner-Moyal transforms of $A(x_f,r_f)$ and $B(x_i,r_i)$
according to
\begin{equation} 
\tilde{A}(p_f,q_f)=\int dx_f \exp\left(-{i\over\hbar}x_f p_f\right) A(x_f,q_f),
\end{equation}
equilibrium correlations may be written as a
double phase-space integral
\begin{equation}
\langle A(t) B\rangle=\int dp_fdq_fdp_idq_i \tilde{A}(p_f,q_f) 
\tilde{B}(p_i,q_i) \tilde{P}(p_f,q_f,t,p_i,q_i).
\end{equation}
This means that we can view $\tilde{P}(p_f,q_f,t,p_i,q_i)$ as a 
quantum mechanical generalization of the classical joint
probability. The Wigner form of the master equation (\ref{evol}), 
which is again a generalized Fokker-Planck equation, was first derived
by Schramm, Jung, and Grabert \cite{schramm} on the basis of 
phenomenological reasoning. Note, that the time-dependent
coefficients are complex. However, in the classical limit
they become real, since the imaginary part of $C(t)$ vanishes, and the
generalized Fokker-Planck equation reduces again to the Adelman
equation.

\subsection{Factorizing initial preparation}
In earlier work it has been frequently assumed \cite{reibold,cal,FV,hu}
that the initial density matrix $W_0$ of the entire system factorizes
according to 
\begin{equation}
W_0=\rho_0 W_R
\end{equation}
where $\rho_0$ is the density matrix of the oscillator, while
$W_R=Z_R\exp(-\beta H_R)$ is the canonical density matrix of the
unperturbed heat bath. Within our approach this situation
cannot be described by a special form of the preparation function
$\lambda(q_i,q_i',\bar{q},\bar{q}')$, however, factorizing initial states
are easily gained by disregarding contributions coming from the
imaginary time path integral (see \cite{report} for details).
The time evolution of the density matrix is then given by
\begin{equation}
\rho(x_f,r_f,t)=\int dr_i dx_i J_{\rm FV}(x_f,r_f,t,x_i,r_i)\rho(x_i,r_i,0),
\label{propFV}
\end{equation}
where
\begin{equation}
J_{\rm FV}(x_f,r_r,t,x_i,r_i)={M|f_3(t)|\over 2\pi\hbar}\exp[{i\over\hbar}
\Sigma_{\rm FV}(x_f,r_r,t,x_i,r_i)]
\label{sigFV}
\end{equation}
and
\begin{eqnarray}
\Sigma_{\rm FV}(x_f,r_f,t,x_i,r_i)&=&
M[x_f r_f f_1(t)+x_i r_i f_2(t)- x_i r_f f_3(t) - x_f r_i f_4(t)]
\nonumber\\
& & + {i\over 2}M\left[x_i^2 R_{\rm FV}^{++}(t)+2x_f x_i R_{\rm FV}^{+-}(t)
+x_f^2 R_{\rm FV}^{--}(t)\right]\label{SigmaFV}.
\end{eqnarray}
The index ``FV'' refers to Feynman and Vernon \cite{FV}. Note that the
definitions of $R^{\pm\pm}(t)$ are now modified
\begin{eqnarray}
R_{\rm FV}^{++}(t)&=&f_3^2(t)K_{q}(t)\nonumber\\
R_{\rm FV}^{+-}(t)&=&f_3(t)[{1\over 2}\dot{K}_{q}(t)-f_1(t) K_{q}(t)]\\
R_{\rm FV}^{--}(t)&=&K_{p}(t)-f_1(t) \dot{K}_{q}(t)+f_1^2(t)
K_{q}(t),\nonumber
\end{eqnarray}
where the $f_i(t)$ are given by (\ref{f12})-(\ref{f4}) and
\begin{eqnarray}
K_{q}(t)&=&{1\over M}\int_0^t ds \int_0^t du K'(s-u)G(s)G(u)\nonumber\\
K_{p}(t)&=&{1\over M}\int_0^t ds \int_0^t du K'(s-u)\dot{G}(s)\dot{G}(u).
\label{KK}
\end{eqnarray}
The function $K'(t)$ is the real part of the real time influence
kernel. Its Laplace transform is related to the Laplace transform of
the damping kernel $\hat{\gamma}(z)$ by \cite{report}
\begin{equation}
\hat{K}'(z)={M\over\hbar\beta}\sum_{n=-\infty}^\infty{z\over z^2-\nu_n^2}
[z\hat{\gamma}(z)-|\nu_n|\hat{\gamma}(|\nu_n|)].\label{hatK}
\end{equation}
The integrals in (\ref{KK}) cannot be expressed in terms of $S(t)$ and
$G(t)$ as it is the case for $R^{\pm\pm}$(t).
The quadratic ansatz (\ref{matr}) now leads to equations
(\ref{dgl1})-(\ref{dgl7}) with the functions $R^{\pm\pm}(t)$ replaced
by $R^{\pm\pm}_{\rm FV}(t)$. There are no equations replacing
(\ref{c1+})-(\ref{c2-}) since oscillator and bath are uncorrelated in
the initial state. This set allows for a solution by the time-dependent 
parameters
\begin{eqnarray}
\gamma_{q}(t)&=&{\ddot{G}^2(t)-\dot{G}(t)\overdots{G}(t)\over 
   \dot{G}^2(t)-G(t)\ddot{G}(t)}\nonumber\\
\gamma_{p}(t)&=&{G(t) \overdots{G}(t)-\dot{G}(t)\ddot{G}(t)\over 
   \dot{G}^2(t)-G(t)\ddot{G}(t)}\nonumber\\
D_{q}(t)&=&{\hbar\over M}\left[
   {1\over 2}\ddot{K}_{q}(t)-K_{p}(t)+\gamma_{q}(t)K_{q}(t)
   +{\gamma_{p}(t)\over 2}\dot{K}_{q}(t)\right]\label{FVerg}\\
D_{p}(t)&=&{\hbar\over M}\left[
   {1\over 2}\dot{K}_{p}(t)+{\gamma_{q}(t)\over 2}\dot{K}_{q}(t)+
   \gamma_{p}(t)K_{p}(t)\right],\nonumber
\end{eqnarray}
The resulting master equation is equivalent to the result by Haake and
Reibold \cite{reibold,anmerk}
who derived it directly from microscopic dynamics. Their formulas for
$K_{q}(t)$ and $K_{p}(t)$ contain frequency integrals that may be
evaluated to obtain (\ref{KK}). Later, this equation was rederived by
Hu, Paz and Zhang \cite{hu} from the path integral representation. The
equivalence can most easily be seen using the simplified derivation
given by Paz \cite{paz}.

We mention that for factorizing initial conditions
the classical limit does not yield the Adelman equation \cite{bmk2}.
The generalized Fokker-Planck operator differs by terms that 
persist over times of the order of the relaxation time.
This means that switching on the interaction with the bath at $t=0$
pathologically affects also the long time behavior of the
system. Usually, the oscillator and the bath are integral
parts of the same system and the factorization assumption is not
appropriate.

\section{Liouville operators for particular types of damping}
So far we have searched for exact master equations. Let us now turn to
the question whether for particular types of damping the dynamics may
be described in terms of approximate Liouville operators valid for
arbitrary preparation functions. Thus, we have to find circumstances
under which $S(t)$ and $G(t)$ take the forms (\ref{ohmic1}) and
(\ref{ohmic2}). Using (\ref{coeffs}) we see that in this case the
Liouville operator (\ref{master}) is time independent and the
coefficients $D_{p}(t)=D_{p}$ and $D_{q}(t)=D_{q}$ read 
\begin{eqnarray}
D_{q}&=&\gamma_{q} \langle q^2\rangle-
{\langle p^2 \rangle\over M^2}\label{erg}\\
D_{p}&=&\gamma_{p}{\langle p^2 \rangle\over M^2}.\nonumber
\end{eqnarray}
Note that this result is independent of the coefficients
$c_1,c_2,d_1,d_2$ in (\ref{ohmic1}) and (\ref{ohmic2}).
The Wigner transform of the density matrix then obeys equation
(\ref{wigner2}) with time independent coefficients $\gamma_{q}$ and
$\gamma_{p}$. This result is in
accordance with the findings of Talkner \cite{talk} on
the most general form of a Liouville operator in Wigner form
compatible with the correct equilibrium expectation values.

To see explicitly when conditions (\ref{ohmic1}) and
(\ref{ohmic2}) hold, we first investigate the consequences of (\ref{ohmic2}).
By virtue of Ehrenfest's theorem $G(t)$ is purely classical
\cite{gwt}, since it is related to the response function
\begin{equation}
\chi(t)=\theta(t) {1\over M} G(t)
\end{equation}
describing the mean non-equilibrium displacement in response to an
applied force. Hence (\ref{ohmic2}) implies that the
classical equation of motion is solved by a sum of two
exponentials. This means essentially Ohmic damping.

\subsection{Ohmic Damping}
For strictly Ohmic damping $\hat{\gamma}(z)=\gamma$ and condition
(\ref{ohmic2}) holds exactly. We have
\begin{equation}
G(t)=c_1 e^{-\lambda_1 t}+c_2 e^{-\lambda_2 t}
\end{equation}
where $\lambda_{1/2}=\gamma/2\pm i\sqrt{\omega_0^2-\gamma^2/4}$ and
$c_{1/2}=\mp (\lambda_1-\lambda_2)^{-1}$.
Thus $ \gamma_{p}=\gamma$ and $\gamma_{q}=\omega_0^2$. We still
have to examine whether $S(t)$ fulfills (\ref{ohmic1}). 
To this purpose it is useful to note that (\ref{defSG}) gives for the 
Fourier transforms
\begin{equation}
\tilde{S}(\omega)={i\hbar\over 2 M}\coth\left({\omega\pi\over\nu}\right)
\tilde{G}(\omega)\label{FDT}
\end{equation}
where
\begin{eqnarray}
\tilde{S}(\omega)&=&\hat{S}(-i\omega)+\hat{S}(i\omega),\nonumber\\
\tilde{G}(\omega)&=&\hat{G}(-i\omega)-\hat{G}(i\omega).
\end{eqnarray}
The latter relations follow from the fact that $S(t)$ is symmetric and
$G(t)$ anti-symmetric. Of course, (\ref{FDT}) is just the familiar 
fluctuation-dissipation theorem. Now, performing the inverse Fourier
transform we find \cite{gwt} 
\begin{equation}
S(t)=d_1 e^{-\lambda_1 t}+d_2 e^{-\lambda_2 t}-\Gamma(t),
\label{Sohm}
\end{equation}
where
\begin{eqnarray}
d_{1/2}&=&c_{1/2}{\hbar\over 2M}
\cot\left({\pi\lambda_{1/2}\over\nu}\right),\nonumber\\
\Gamma(t)&=&{2\gamma\over M\beta}\sum_{n=1}^\infty
{\nu_n\exp(-\nu_n t)\over(\omega_0^2+\nu_n^2)^2-\gamma^2\nu_n^2}.
\end{eqnarray}
This means that we have to find conditions under which $\Gamma(t)$
may be disregarded. Now $\Gamma(t)$ decays at least as $\exp(-\nu t)$,
where $\nu\equiv\nu_1=2\pi/k_B T$. Therefore, for temperatures
$T\gg\hbar\gamma/4\pi k_B$, we have $\nu\gg {\rm Re}(\lambda_1), 
{\rm Re}(\lambda_2)$ and $\Gamma(t)$ decays faster than the first two
terms in (\ref{Sohm}). Hence for $t\gg\nu^{-1}$ the last term in
(\ref{Sohm}) may be disregarded and $S(t)$ is of the form
(\ref{ohmic2}).
However, in the strictly Ohmic case we do not have a well-defined
Liouville operator since the sum (\ref{defpq}) for $\langle
p^2\rangle$ is logarithmically divergent leading to associated
divergences of the coefficients $D_{q}$ and $D_{p}$ in (\ref{erg}).
To avoid this divergence we have to take the high frequency behavior
of the damping coefficient into account which implies in realistic
cases
\begin{equation}
\lim_{z\to\infty}\hat{\gamma}(z)=0
\end{equation}
\subsection{Drude Regularization}
A more realistic behavior of the damping coefficient is modeled by
\begin{equation}
\hat{\gamma}(z)={\gamma \omega_D\over z+\omega_D},
\end{equation}
often referred to as Drude damping. Using (\ref{defSG}) we then find
\begin{equation}
G(t)=c_1 e^{-\lambda_1 t}+c_2 e^{-\lambda_1 t}+c_3 e^{-\lambda_3 t}
\label{GDrude}
\end{equation}
where 
\begin{equation}
\lambda_{1/2}=\alpha\pm i\eta,\quad \lambda_3=\delta
\label{roots}
\end{equation}
are the solutions of
\begin{equation}
z^3-\omega_D z^2+(\omega_0^2+\gamma\omega_D)z-\omega_0^2\omega_D=0
\label{kubisch}
\end{equation}
and
\begin{eqnarray}
c_1&=&-{i\over 2\eta}{\alpha-i\eta+\delta\over \alpha
+i\eta-\delta}\label{c1},\nonumber\\ 
c_2&=&{i\over 2\eta}{\alpha+i\eta+\delta\over \alpha
-i\eta-\delta},\label{c2}\\ 
c_3&=&{2\alpha\over (\alpha-i\eta-\delta)(\alpha+i\eta-\delta)}.\nonumber
\end{eqnarray}
This result holds also in the overdamped case where $\eta$
becomes imaginary.
To calculate $S(t)$ we have to evaluate the inverse Fourier transform
of (\ref{FDT}) by contour integration. Using (\ref{GDrude}) one finds
\cite{anker}
\begin{equation}
S(t)=d_1 e^{-\lambda_1 t}+d_2 e^{-\lambda_1 t}+d_3 e^{-\lambda_3
t}-\Gamma(t)\label{Sdrude}
\end{equation}
where
\begin{eqnarray}
d_i&=& c_i{\hbar\over 2 M}\cot\left({\pi\lambda_i\over\nu}\right),
\quad i=1,2,3\label{ddd}\\
\Gamma(t)&=&{2\gamma\over M\beta}
\sum_{n=1}^\infty {\omega_D^2\nu_n e^{-\nu_n t}\over
(\lambda_1^2-\nu_n^2)(\lambda_2^2-\nu_n^2)(\lambda_3^2-\nu_n^2)}.
\end{eqnarray}
From these results we see that $S(t)$ and $G(t)$ are of the forms
(\ref{ohmic1}) and (\ref{ohmic2}), respectively, if the exponentials
$\exp(-\lambda_3 t)$ and $\exp(-\nu_n t)$ decay much faster than 
$\exp(-\lambda_{1/2}t)$.
This is the case for 
\begin{equation}
{\rm Re}(\lambda_1), {\rm Re}(\lambda_2)\ll\delta,\ \nu.
\end{equation}
Using the Vieta relations
\begin{eqnarray}
2\alpha+\delta&=&\omega_D\nonumber\\
\alpha^2+\eta^2&=&\omega_0^2\omega_D/\delta\label{vieta}\\
\alpha^2+\eta^2+2\alpha\delta&=&\omega_0^2+\gamma\omega_D,\nonumber
\end{eqnarray}
we find that ${\rm Re}(\lambda_{1/2})\ll\delta$ implies 
\begin{equation}
\alpha\ll\delta. \label{aldel}
\end{equation}
Further, the relations (\ref{vieta}) yield
\begin{equation}
{\gamma\over\omega_D}=2{\alpha\over\delta}{1+(\omega_0/\delta)^2\over 
(1+2\alpha/\delta)^2}.
\end{equation}
In view of (\ref{aldel}) this gives \cite{bem3}
\begin{equation}
\gamma/\omega_D = 2 \alpha/\delta-8(\alpha/\delta)^2
+{\cal O}((\alpha/\delta)^3).
\end{equation}
Hence, it is natural to use $\gamma/\omega_D$ as small parameter to
determine the roots of (\ref{kubisch}). Up to first order in
$\gamma/\omega_D$ one obtains for the roots (\ref{roots})
\begin{eqnarray}
\alpha&=&{\gamma\over 2}{\omega_D^2\over\omega_D^2+\omega_0^2}\nonumber\\
\eta&=&\sqrt{\omega_0^2+2\alpha\omega_0^2/\omega_D-\alpha^2}\\
\delta&=&\omega_D-2\alpha.\nonumber
\end{eqnarray}
These relations are valid for arbitrary ratios of $\gamma$ and $\omega_0$.
Also $\omega_D/\omega_0$ is not necessarily large.
As a consequence of this analysis we find that only under the conditions
\begin{equation}
\gamma\ll\omega_D,\nu\label{god}
\end{equation}
and 
\begin{equation}
t\gg\omega_D^{-1},\ \nu^{-1}.
\end{equation}
$S(t)$ may be approximated by the first two terms in (\ref{Sdrude}).
Hence, for sufficiently large Drude cutoff and sufficiently high
temperatures,  $k_B T\gg\hbar\gamma$,
the oscillator dynamics can be described by an approximate Liouville
operator with the coefficients
\begin{eqnarray}
\gamma_{q}&=&\alpha^2+\eta^2\nonumber\\
\gamma_{p}&=&2\alpha.
\end{eqnarray}
This combines with (\ref{erg}) to yield the approximate master equation
\begin{eqnarray}
\dot{\rho}(t)&=&-{iM\over\hbar}(\alpha^2+\eta^2)\left[q,{1\over 2}\{q,\rho(t)\}
   +{i\over\hbar}\langle q^2\rangle[p,\rho(t)]\right]\nonumber\\
& &-{2i\over\hbar}\alpha\left[q,{1\over 2}\{p,\rho(t)\}
   -{i\over\hbar}\langle p^2\rangle[q,\rho(t)]\right]\label{appmeq}\\
& &-{i\over M\hbar}\left[p,{1\over 2}\{p,\rho(t)\}
   -{i\over\hbar}\langle p^2\rangle[q,\rho(t)]\right]\nonumber.
\end{eqnarray}
first derived by Haake and Reibold \cite{reibold}.

The equilibrium variances can be calculated analytically \cite{gwt}.
In the strictly Ohmic limit $\omega_D\to\infty$, $\langle
q^2\rangle$ is a regular expression but $\langle p^2\rangle$ diverges
logarithmically. If we disregard terms of the order
$\omega_0/\omega_D$, $\gamma/\omega_D$, $\nu/\omega_D$, the divergent
part of $\langle p^2\rangle$ is given by
\[
{M\hbar\gamma\over\pi}\ln\left({\omega_D\over\nu}\right).
\]
Hence, for the classical limit it is not sufficient to have
$\nu\gg\omega_0$, that is $k_B T\gg\hbar\omega_0$, rather we also need
$\gamma\ln(\omega_D/\nu)\ll\nu$. Thus the strictly Ohmic limit
$\omega_D\to\infty$ can only be taken after the classical limit. 
With this sequence of limits we obtain the classical coefficients
\begin{equation}
\gamma_{q}=\omega_0^2,\  \gamma_{p}=\gamma,\ D_{q} = 0,\ 
D_{p} = {\gamma k_B T\over M}.
\label{ohmsch}
\end{equation}
The associated Liouville operator is equivalent to the classical
Fokker-Planck operator of the Kramers equation \cite{kramers}.
We stress again that the results in this section remain valid for
strong damping provided (\ref{god}) is satisfied.

\subsection{Weak damping with arbitrary frequency dependence}
In this section we show that $S(t)$ and $G(t)$ always take the forms
(\ref{ohmic1}) and (\ref{ohmic2}) in the limit of weak damping. 
Let us assume that the 
damping kernel $\gamma(t)$ has a high frequency cutoff $\omega_c$ and
that its Laplace transform is an analytic function in the vicinity of
$-i\omega_0$. We introduce a typical damping strength by
\begin{equation}
\gamma_c=\int_0^\infty ds \gamma(s)\cos(\omega_0 s).
\end{equation}
Then the weak damping condition is
\begin{equation}
\gamma_c\ll\omega_0,\omega_c,\nu.
\end{equation}
Apart from this $\gamma(t)$ is not assumed to have additional properties. 

In the limit considered one can determine the poles of $\hat{G}(z)$
from (\ref{defSG}) perturbatively. 
To first order in $\gamma_c$ the poles are 
\begin{equation}
\lambda_{1/2}={\gamma_c\over 2}\pm i\left(\omega_0+{\gamma_s\over 2}\right)
\end{equation}
where 
\begin{equation}
\gamma_c+i\gamma_s=\hat{\gamma}(-i\omega_0).
\end{equation}
Performing the inverse Laplace transform of (\ref{defSG}) we find 
\begin{equation}
G(t)={i\over 2\omega_0}\left(e^{-\lambda_1 t}-e^{-\lambda_2 t}\right)
\end{equation}
with residues in zeroth order. Higher order corrections would
depend on the specific form of $\hat{\gamma}(z)$ but need not be
determined because the diffusion constants (\ref{erg}) are independent
of the residues. Thus, in the weak damping limit $G(t)$ is of the form
(\ref{ohmic2}). From (\ref{FDT}) we see that $\tilde{S}(\omega)$ has the same
poles as $\tilde{G}(\omega)$ and, in addition, poles at $i\nu_n$ 
($n$ integer). 
As shown above, we can disregard the terms
coming from the poles at $i\nu_n$ for times greater than the thermal
relaxation time $\nu^{-1}$. Thus, $S(t)$ is effectively of the form
(\ref{ohmic1}).
Inserting $\lambda_{1/2}$ into (\ref{erg}), we find to leading order
in the damping strength
\begin{eqnarray}
D_{q}&=& {1\over M\beta}\sum_{n=-\infty}^\infty {\gamma_s\omega_0-|\nu_n|
\hat{\gamma}(|\nu_n|)\over \omega_0^2+\nu_n^2}\nonumber\\
D_{p}&=& {1\over M\beta}\sum_{n=-\infty}^\infty {\gamma_c\omega_0^2\over
\omega_0^2+\nu_n^2}=\gamma_c{\hbar\omega_0\over
2M}\coth\left({\omega_0\pi\over\nu}\right).
\end{eqnarray}
This result can be expressed in terms of the Laplace
transform (\ref{hatK}) of the real part of the influence kernel
\begin{equation}
D_{p}+ i\omega_0 D_{q}={\hbar\over M^2}\hat{K}'(-i\omega_0).
\end{equation}
Thus, $D_q$ and $D_p$ are essentially given by the sine- and
cosine-moments of $K'(t)$. With $K_c+iK_s=\hat{K}'(-i\omega_0)$ the
master equation takes the form
\begin{eqnarray}
\dot{\rho}(t)&=&-{i\over\hbar}[{p^2\over 2M}+{M
(\omega_0^2+\omega_0\gamma_s)\over 2}q^2,\rho(t)]
-{i\gamma_c\over 2\hbar}[q,\{p,\rho(t)\}]\nonumber\\
&&-{K_s\over M\hbar\omega_0}[p,[q,\rho(t)]]
-{K_c\over\hbar}[q,[q,\rho(t)]].\label{Agener}
\end{eqnarray}
This general weak coupling master equation is given in terms of four
dissipation coefficients. $\gamma_s$ leads to a frequency shift and
may be absorbed by renormalizing $\omega_0$. $\gamma_c$ is the
classical damping coefficient. The coefficients $K_c$ and $K_s$ depend
on the temperature. While $K_c$ equals $M\gamma_c/\hbar$ times the
average energy of a quantum oscillator of frequency $\omega_0$, $K_s$
depends on the specific form of $\hat{\gamma}(z)$ and can be
calculated analytically only in certain cases. One of these is the Drude model.
Then the moments $\gamma_{c/s}$ and $K_{c/s}$ are
readily evaluated to read
\begin{eqnarray}
\gamma_c&=&{\gamma\omega_D^2      \over \omega_D^2+\omega_0^2}\nonumber\\
\gamma_s&=&{\gamma\omega_0\omega_D\over \omega_D^2+\omega_0^2}\nonumber\\
K_c&=&{\gamma_c\omega_0 M\over 2}\coth\left({\omega_0\pi\over\nu}\right)
\label{ggKK}\\
K_s&=&{\omega_0 M\over\pi}\left\{{\nu\gamma_s\over 2\omega_0}+\gamma_c
{\rm Re}\left[
\psi(1+i{\omega_0\over\nu})-\psi(1+{\omega_D\over\nu})\right]\right\}
\nonumber
\end{eqnarray}
In the strictly Ohmic limit $\omega_D\to\infty$ we have
$\gamma_c=\gamma$, $\gamma_s=0$ but
\begin{equation}
K_s=-{\gamma\omega_0 M\over\pi}\ln\left({\omega_D\over\nu}\right)
\end{equation}
is logarithmically divergent.
The master equation (\ref{Agener}) with the coefficients (\ref{ggKK})
can be compared with the well-known Agarwal equation \cite{agar}
\begin{equation}
\dot{\rho}(t)=-{i\over\hbar}[{p^2\over 2M}+{M
\omega_0^2\over 2}q^2,\rho(t)]
-{i\kappa\over \hbar}[q,\{p,\rho(t)\}]
-\kappa{M\omega_0\over\hbar}\coth\left({\omega_0\pi\over\nu}\right)
[q,[q,\rho(t)]],\label{agarwaleq}
\end{equation}
which was derived with the help of
projection operator techniques from the same
microscopic model using the Born approximation in conjunction
with a shot-memory approximation. As a main difference, we see that in
Agarwal's equation the $K_s$ term is absent. This term is only negligible if 
\begin{equation}
\omega_0\ll\omega_D\ll\nu.
\end{equation}
Hence, the new
master equation (\ref{Agener}) is a generalization of the Agarwal
equation.
\subsection{Connection to Lindblad theory}
The approximate time-independent Liouville operators studied above
describe the dynamics after the decay of fast transients. 
Markovian Liouville operators like these are often discussed in the
context of Lindblad theory \cite{lind}. This theory establishes the most
general form of generators ${\cal L}$ of dissipative quantum dynamics
$\dot{\rho}(t)={\cal L}\rho(t)$ preserving the positivity of density
operators. The Lindblad master equation reads
\begin{equation}
\dot{\rho}(t)=
-{i\over\hbar} [\tilde{H},\rho(t)]+{1\over 2\hbar}\sum_i [L_i\rho(t),L_i^+]+
[L_i,\rho(t) L_i^+]\label{lind}
\end{equation}
where $L_l$ are arbitrary operators and $\tilde{H}$ is a Hermitian operator.
Using results by Sandulescu and Scutaru \cite{sand} it is easily seen
that all the above derived time-independent Liouville operators are
not of Lindblad form. This is not too astonishing since the master
equations derived hold only for times $t>t_0$ where $t_0$ is larger
than an inverse cutoff frequency and $\nu^{-1}$. The short time dynamics for
$t\lesssim t_0$ reduces the density matrix to a subspace where the
fast components have decayed. The Markovian master equation holds
within this subspace only, while Lindblad theory requires validity for
{\em any} reduced density matrix. This is of course not necessary as
has been emphasized again recently \cite{pechukas,gnutzmann}.

However, we will show that in the weak coupling limit further
coarse-graining will result in a Lindblad operator.
To this aim we first write the weak-coupling master equation 
(\ref{Agener}) in the form
\begin{equation}
\dot{\rho}(t)={\cal L}\rho(t)={\cal L}_0\rho(t)+\gamma {\cal L}_1\rho(t).
\end{equation}
In terms of the usual creation and annihilation operators
$a^\dagger$, $a$ we have
\begin{equation}
{\cal L}_0=-i\omega_0[a^\dagger a,\cdot\ ].
\end{equation}
Using the operators
\begin{equation}
{\cal P}_n=\sum_k |k\rangle\langle k+n|,
\end{equation}
where $|k\rangle$ are the eigenstates of $a^\dagger a$,
${\cal L}_0$ may be written as
\begin{equation}
{\cal L}_0=\sum_n i\omega_0 n {\cal P}_n.\label{L0P}
\end{equation}
Further
\begin{eqnarray}
\gamma {\cal L}_1&=&-i{\gamma_s\over 2}[a^\dagger a,\cdot\ ]
+{\gamma_c-i\gamma_s\over 4}[{a^\dagger}^2,\cdot\ ]
-{\gamma_c+i\gamma_s\over 4}[a^2,\cdot\ ]\nonumber\\
&&+\gamma_\downarrow \left([a\ \cdot\ ,a^\dagger]+[a,\cdot\ a^\dagger]\right)
+\gamma_\uparrow \left([a^\dagger\ \cdot\ ,a]+[a^\dagger,\cdot\ a]\right)\\
&&+{K_c+iK_s\over 2M\omega_0}\left([a^\dagger\ \cdot\ ,a^\dagger]
+[a^\dagger,\cdot\ a^\dagger]\right)
+{K_c-iK_s\over 2M\omega_0}\left([a\ \cdot\ ,a]+[a,\cdot\ a]\right)\nonumber,
\end{eqnarray}
where we have introduced
\begin{equation}
\gamma_{\downarrow\uparrow}={K_c\over 2M\omega_0}\pm{\gamma_c\over 4}
={\gamma_c\over
4}\left[\coth\left({\omega_0\pi\over\nu}\right)\pm 1\right].
\end{equation}
The time evolution is formally given by
\begin{equation}
\rho(t)=e^{{\cal L} t}\rho(0).
\end{equation}
Now we rewrite this by a well-known operator identity
\begin{equation}
\rho(t)=e^{{\cal L}_0t}\rho(0)+\gamma\int_0^t ds e^{{\cal L}_0(t-s)}{\cal L}_1e^{{\cal L} s}\rho(0).
\label{pert}
\end{equation}
For weak damping and times $t\ll\gamma^{-1}$ the operator $e^{{\cal L}s}$ in
the integrand may be replaced by $e^{{\cal L}_0s}$.
Inserting then (\ref{L0P}) into (\ref{pert}) we find
\begin{equation}
\rho(t)=e^{{\cal L}_0 t}\left[1+\gamma\sum_{n,m}\int_0^t ds e^{i\omega_0(n-m)s}
{\cal P}_n{\cal L}_1{\cal P}_m\right]\rho(0)
\label{pert1}
\end{equation}
Further coarse-graining is achieved by demanding
\begin{equation}
t\gg\omega_0^{-1}.
\end{equation}
Then, by performing the time integral, the
off-diagonal terms ($n\neq m$) are seen to be smaller than the
diagonal terms by a factor $(\omega_0t)^{-1}$. This means that in the
time window $\omega_0^{-1}\ll t\ll\gamma^{-1}$
\begin{equation}
\rho(t)=e^{{\cal L}_0t}\left[1+\gamma t\tilde{{\cal L}}_1\right]\rho(0)
\label{window}
\end{equation}
where we have introduced the {\em effective} dissipative Liouville
operator
\begin{equation}
\tilde{{\cal L}}_1=\sum_n {\cal P}_n{\cal L}_1{\cal P}_n.
\end{equation}
The density matrix (\ref{window}) coincides with the solution of 
the master equation
\begin{equation}
\dot{\rho}(t)=({\cal L}_0+\gamma \tilde{{\cal L}}_1)\rho(t)
\end{equation}
for
\begin{equation}
t\ll\gamma^{-1}.
\end{equation}
Thus, within the time-window $\omega_0^{-1}\ll t \ll\gamma^{-1}$ the two
operators ${\cal L}_0+\gamma {\cal L}_1$ and ${\cal L}_0+\gamma \tilde{{\cal L}}_1$ give the same
dynamics.
The operator $\tilde{{\cal L}}_1$ may be evaluated further. It is seen
that only ``non-rotating'' terms containing equal numbers of creation and
annihilation operators survive the coarse-graining in time. 
The resulting master equation
\begin{equation}
\dot{\rho}(t)=-i(\omega_0+{\gamma_s\over 2})[a^+a,\rho(t)]+\gamma_\uparrow
\left([a^+\rho(t),a]+[a^+,\rho(t) a]\right)+\gamma_\downarrow
\left([a\rho(t),a^+]+[a,\rho(t) a^+]\right)
\end{equation}
was first derived by Weidlich and Haake \cite{hw}
from a microscopic model for the damped motion of a single mode of the
electromagnetic field in a cavity. The generator defined by this
master equation is of Lindblad form. However, the resulting mean value
equations violate Ehrenfest's theorem, in particular 
\begin{equation}
{\partial \over \partial t}\langle q(t)\rangle\neq \langle p(t)\rangle/M.
\end{equation}
This is due to the fact that on the coarse-grained time scale $\Delta
t\gg\omega_0^{-1}$ the variables $p(t)/M\omega_0$ and $q(t)$ exchange identity
frequently and only a time averaged version of the mean value equations
must be obeyed.
\section{conclusions}
Based on results of the path integral technique we have examined
quantum master equations for the damped harmonic oscillator. A new
exact generalized master equation describing the relaxation of initial
thermal conditions was derived. This equation was shown to be a
quantum mechanical generalization of Adelman's Fokker-Planck equation.
We also have given an exact Liouville operator describing the time
evolution of equilibrium correlation functions which likewise reduces
to the Adelman Fokker-Planck operator in the classical limit. The
fact, that two different quantum generalizations of the Adelman
operator must be used for the relaxation of expectation values and the
regression of fluctuations is intimately connected with the failure of
the Onsager regression hypothesis in the quantum regime. Indeed, the two
Liouville operators (\ref{neu}) and (\ref{Liou}) are only identical
if $S(t)$ is proportional to $G(t)$.

Apart from these exact results we have studied in detail the range of
parameters leading to quantum master equations with time-independent
generator. In the case of strong damping a
time-independent Liouville operator is obtained approximately for
essentially frequency-independent damping. However,
strictly Ohmic damping is ill-behaved in the quantum case, and
the appropriate generalization of the classical Fokker-Planck 
process is given by a low-frequency Ohmic model with high frequency
cutoff such as the Drude model. This is not too
amazing since already in the classical limit the Adelman operator
becomes time-independent only for frequency-independent damping.

On the
other hand, in the case of weak damping the detailed frequency
dependence of the damping coefficient is unimportant. We have derived a new
generalized master equation valid for arbitrary
weak damping as long as the sine and cosine moments of the damping
kernel exist. The new weak-coupling master equation
is more general than the well-known Agarwal equation.
We have explained why the Liouville operator is not of Lindblad form.
However, time coarse-graining leads to a
generator of Lindblad form. The resulting coarse-grained master
equation was found to be the Weidlich-Haake equation also known as
the quantum optical master equation. Due to the time coarse-graining
only a time-averaged version of the mean value equation is obeyed
leading to an apparent contradiction with the Ehrenfest theorem. 

In summary, we have derived several new generalized master equations
for the damped quantum oscillator for various cases of interest. 
In view of the new results earlier findings were put in proper perspective.

\section{acknowledgments}
The authors would like to thank J. Ankerhold, J. Hainz, and
F. J. Weiper for valuable discussions and S. Gnutzmann and F. Haake
for communicating their results prior to publication. Financial
support was provided by the Deutsche Forschungsgemeinschaft through
the Sonderforschungsbereich 276.


\begin{thebibliography}{99}
\bibitem{GtoG+}{In Ref. \cite{report} $G(t)$ is denoted by $G_+(t)$}
\bibitem{anmerk} There are some typos in equation ($9.4"$) of
        Ref. \cite{reibold}: $\dot{A}A$ must be replaced by
        $\dot{A}\overdots{A}$, $A^2$ by $A\overdots{A}$ and the sign in front
        of $\ddot{X}$ must be changed.
\bibitem{bem3} We assume that $\omega_D/\omega_0$ is of order 1 
        or larger.
\end{thebibliography}
\end{document}